\documentclass{aa}
\usepackage{graphicx}
\usepackage{color}
\usepackage{subfig}
\usepackage{float}
\usepackage{amsmath}

\begin{document}
   \title{Magnetic field structure in single late-type giants: $\beta$ Ceti in 2010 -- 2012\thanks{Based on observations obtained at the Bernard Lyot T\'elescope (TBL, Pic du Midi, France) of the Midi-Pyr\'en\'ees Observatory, which is operated by the Institut National des Sciences de l'Univers of the Centre National de la Recherche Scientifique of France and Universit\'e de Toulouse, and at the Canada-France-Hawaii Telescope (CFHT), which is operated by the National Research Council of Canada, the Institut National des Sciences de l'Univers of the Centre National de la Recherche Scientifique of France, and the University of Hawaii.}}


   \author{S. Tsvetkova\inst{1}, P. Petit\inst{2,3}, M. Auri\`ere\inst{2,3}, R. Konstantinova-Antova\inst{1,3}, G.A. Wade\inst{4}, C. Charbonnel\inst{5,2}, T. Decressin\inst{5}, R. Bogdanovski\inst{1}}

   \offprints{S. Tsvetkova, stsvetkova@astro.bas.bg}

   \institute{Institute of Astronomy and NAO, Bulgarian Academy of Sciences, 72 Tsarigradsko shose, 1784 Sofia, Bulgaria\\
              \email{stsvetkova@astro.bas.bg}
     \and CNRS, Institut de Recherche en Astrophysique et Plan\'etologie, IRAP, 14 Avenue Edouard Belin, 31400 Toulouse, France
     \and Universit\'e de Toulouse, UPS-OMP, IRAP, Toulouse, France
     \and Department of Physics, Royal Military College of Canada, PO Box 17000, Station `Forces', Kingston, Ontario, Canada K7K 4B4
     \and Geneva Observatory, University of Geneva, 51, Chemin des Maillettes, 1290 Versoix, Switzerland\\
              }

   \date{Received ; }


  \abstract
   {}
   {We study the behavior of the magnetic field and the line activity indicators of the single late-type giant $\beta$ Ceti. Using spectropolarimetric data, we aim to reconstruct the magnetic field structure on the star's surface and to present the first magnetic maps for $\beta$ Ceti.}
   {The data were obtained using two spectropolarimeters -- Narval at the Bernard Lyot T\'elescope, Pic du Midi, France, and ESPaDOnS at CFHT, Hawaii. Thirty-eight circularly-polarized spectra have been collected in the period June 2010 -- January 2012. The least square deconvolution method was applied for extracting high signal-to-noise ratio line profiles, from which we measured the surface-averaged longitudinal magnetic field B$_{l}$. Chromospheric activity indicators CaII K, H$\alpha$, CaII IR (854.2 nm), and radial velocity were simultaneously measured, and their variability was analyzed along with the behavior of B$_{l}$. The Zeeman Doppler imaging (ZDI) inversion technique  was employed for reconstructing the large-scale magnetic field and two magnetic maps of $\beta$ Ceti are presented for two periods (June 2010 -- December 2010 and June 2011 -- January 2012).}
   {The B$_{l}$ stays with a same positive polarity for the whole observational period and shows significant variations in the interval 0.1 -- 8.2~G. The behavior of the line activity indicators is in good agreement with the B$_{l}$ variations. Searching for periodic signals in the Stokes V time series, we found a possible rotation period of 215~days. The two ZDI maps show a mainly axisymmetric and poloidal magnetic topology and a simple surface magnetic field configuration dominated by a dipole. Little evolution is observed between the two maps, in spite of a 1 yr interval between both subsets. We also use state-of-the-art stellar evolution models to constrain the evolutionary status of $\beta$ Ceti. We derive a mass of 3.5~M$_{\odot}$ and propose that this star is already in the central helium-burning phase.}
   {Considering all our results and the evolutionary status of the star, we suggest that dynamo action alone may not be efficient enough to account for the high magnetic activity of $\beta$ Ceti. As an alternate option, we propose that it is a descendant of an Ap star presently undergoing central helium-burning and still exhibiting a remnant of the Ap star magnetic field.}

   \keywords{star: individual:$\beta$ Ceti -- magnetic field -- activity -- dynamo}
   \authorrunning{Tsvetkova et al.}
   \titlerunning{Magnetic field structure in $\beta$ Ceti}
   \maketitle
%

\section{Introduction}

Studying magnetic activity in cool single giants is crucial for understanding the role of magnetic fields in the late stages of stellar evolution. By analogy with cool stars on the main sequence, large-scale magnetic fields of cool evolved stars may be the product of a large-scale dynamo, triggered by the interplay between internal differential rotation and convection in the stellar envelope (Konstantinova-Antova et al. 2010; 2012). Their magnetic fields may also be a fossil remnant inherited from previous evolutionary stages, still frozen in stable internal layers of the star and either pervading convective layers up to the surface or interacting with a global dynamo (Auri\`ere et al. 2008; 2011; 2012).

This second scenario may be of particular importance for those cool giants that are the descendants of Ap/Bp stars. The fossil field hypothesis is generally preferred to account for the strong and topologically simple surface magnetic fields of main sequence Ap/Bp stars, and this fossil magnetism may still remain hidden in the internal layers of Ap star progenies after their departure from the main sequence, leading to magnetic properties different from those of a global dynamo.

A third possible contributor to the surface magnetism of cool giants may be a high tangled magnetic pattern coming from a small-scale dynamo. While local dynamos are probably active in all stars with convective zones, their direct observational signatures may only be detectable in the most evolved objects (cool supergiants), for which the larger pressure scale height results in much larger convective cells (Schwarzschild 1975). This last type of magnetism is presumably responsible for the weak magnetic field detected at the surface of the slowly rotating supergiant Betelgeuse (Auri\`ere et al. 2010, Petit et al. 2013).

Observations are obviously essential for deciding between these different scenarios (that may act independently or together, depending on the object) or unveil other possible physical explanations for the manifestations of atmospheric activity recorded in cool, evolved stars. Recent reviews of different types of active stars, including cool giants, have been presented by Berdyugina (2005) and Donati \& Landstreet (2009). In addition to well-known and somewhat atypical evolved objects (e.g. components of RS CVn systems and FK Com stars), weak surface magnetic fields, ranging from a few gauss to a few tens of gauss, are now detected in more common, single G-K giants (Konstantinova-Antova et al. 2008; 2009; Auri\`ere et al. 2009). Similar detections were also reported for a few M giants, e.g. EK Boo (of M5 spectral type, Konstantinova-Antova et al. 2010). The surface magnetic topology, sometimes along with its temporal evolution, has been reconstructed for a small number of cool giants -- EK Eri, which is a descendant of an Ap star (Auri\`ere et al. 2011), V390 Aur, an actually single giant (although it is a binary component, it can be considered single for our purpose because synchronization plays no role as far as its fast rotation and magnetic activity are concerned; Konstantinova-Antova et al. 2012), the early K-type primary giant of the RS CVn binary IM Peg (Berdyugina \& Marsden 2006), and the FK Com star HD 199178 (Petit et al. 2004).

\begin{table*}[!htc]
\centering
\begin{center}
\caption{Measured longitudinal magnetic field, activity indicators, radial velocity, and their uncertainties for $\beta$ Ceti. In the first column capital letters N and E stand for Narval (TBL, Pic du Midi, France) and ESPaDOnS (CFHT), respectively. The fifth column gives the signal-to-noise ratio (S/N) of each Stokes V LSD profile.}  
\label{table:1}      
\centering                          
\begin{tabular}{c c c c c c c c c c c c c c}        
\hline\hline                 
     & Date&    & Rot.& S/N&        &         &          &         &        &         & B$_{l}$& $\sigma$& RV \\
Inst.&   UT& HJD&phase&(LSD)& CaII K& $\sigma$& H$\alpha$& $\sigma$& CaII IR& $\sigma$&  [G]   &    [G]  & [km/s]\\ 
\hline                        
\\
E & 20 Jun 10 & 2455368.123 & 0.891 & 73~412 & 0.191 & 0.005 & 0.211 & 0.002 & 0.134 & 0.003 & 4.4 & 0.4 & 13.36\\
E & 22 Jun 10 & 2455370.134 & 0.901 & 63~268 & 0.194 & 0.004 & 0.214 & 0.001 & 0.135 & 0.003 & 3.9 & 0.4 & 13.36\\
E & 17 Jul 10 & 2455395.140 & 0.017 & 78~329 & 0.210 & 0.004 & 0.220 & 0.001 & 0.141 & 0.001 & 7.4 & 0.3 & 13.36\\
E & 18 Jul 10 & 2455396.132 & 0.022 & 64~598 & 0.216 & 0.005 & 0.217 & 0.001 & 0.144 & 0.002 & 8.1 & 0.4 & 13.38\\
E & 26 Jul 10 & 2455404.138 & 0.059 & 30~276 & 0.207 & 0.012 & 0.215 & 0.001 & 0.146 & 0.000 & 8.2 & 0.9 & 13.47\\
E & 05 Aug 10 & 2455414.006 & 0.105 & 64~982 & 0.218 & 0.005 & 0.213 & 0.001 & 0.147 & 0.001 & 7.2 & 0.4 & 13.45\\
N & 07 Aug 10 & 2455415.633 & 0.112 & 50~740 & 0.197 & 0.007 & 0.211 & 0.002 & 0.149 & 0.001 & 7.0 & 0.5 & 13.44\\
N & 17 Aug 10 & 2455425.685 & 0.159 & 59~498 & 0.201 & 0.004 & 0.221 & 0.002 & 0.148 & 0.003 & 4.5 & 0.5 & 13.43\\
N & 03 Sep 10 & 2455442.699 & 0.238 & 55~431 & 0.189 & 0.007 & 0.218 & 0.001 & 0.147 & 0.001 & 2.8 & 0.4 & 13.36\\
N & 19 Sep 10 & 2455459.473 & 0.316 & 69~382 & 0.187 & 0.005 & 0.211 & 0.002 & 0.142 & 0.001 & 1.2 & 0.4 & 13.36\\
N & 26 Sep 10 & 2455466.500 & 0.349 & 70~223 & 0.198 & 0.006 & 0.215 & 0.001 & 0.147 & 0.002 & 0.1 & 0.4 & 13.35\\
N & 06 Oct 10 & 2455475.518 & 0.391 & 40~356 & 0.212 & 0.007 & 0.215 & 0.002 & 0.155 & 0.002 & 1.8 & 0.7 & 13.43\\
N & 13 Oct 10 & 2455483.475 & 0.428 & 72~465 & 0.230 & 0.005 & 0.213 & 0.001 & 0.165 & 0.001 & 3.6 & 0.4 & 13.44\\
E & 16 Oct 10 & 2455485.832 & 0.439 & 65~746 & 0.216 & 0.003 & 0.223 & 0.002 & 0.155 & 0.002 & 3.9 & 0.4 & 13.43\\
E & 17 Oct 10 & 2455486.863 & 0.444 & 68~396 & 0.222 & 0.006 & 0.217 & 0.001 & 0.154 & 0.001 & 3.8 & 0.3 & 13.42\\
N & 20 Oct 10 & 2455490.471 & 0.460 & 31~993 & 0.211 & 0.016 & 0.213 & 0.001 & 0.158 & 0.001 & 5.1 & 0.8 & 13.46\\
N & 12 Nov 10 & 2455513.360 & 0.567 & 37~268 & 0.199 & 0.004 & 0.212 & 0.002 & 0.149 & 0.004 & 4.9 & 0.7 & 13.40\\
E & 16 Nov 10 & 2455516.831 & 0.583 & 61~599 & 0.202 & 0.003 & 0.222 & 0.002 & 0.142 & 0.003 & 4.0 & 0.4 & 13.35\\
E & 21 Nov 10 & 2455521.798 & 0.606 & 59~370 & 0.197 & 0.007 & 0.212 & 0.002 & 0.137 & 0.001 & 4.2 & 0.5 & 13.31\\
E & 22 Nov 10 & 2455522.718 & 0.610 & 73~414 & 0.195 & 0.006 & 0.212 & 0.004 & 0.137 & 0.004 & 3.9 & 0.4 & 13.34\\
N & 26 Nov 10 & 2455527.368 & 0.632 & 61~161 & 0.185 & 0.004 & 0.212 & 0.001 & 0.136 & 0.003 & 4.5 & 0.4 & 13.26\\
E & 28 Nov 10 & 2455528.855 & 0.639 & 65~837 & 0.196 & 0.006 & 0.212 & 0.002 & 0.137 & 0.002 & 3.9 & 0.4 & 13.32\\
N & 04 Dec 10 & 2455535.349 & 0.669 & 41~811 & 0.175 & 0.009 & 0.207 & 0.001 & 0.138 & 0.003 & 4.8 & 0.6 & 13.35\\
N & 12 Dec 10 & 2455543.383 & 0.706 & 29~688 & 0.175 & 0.010 & 0.211 & 0.001 & 0.139 & 0.002 & 5.0 & 0.8 & 13.32\\
N & 14 Dec 10 & 2455545.292 & 0.715 & 22~465 & 0.167 & 0.013 & 0.208 & 0.002 & 0.147 & 0.003 & 5.3 & 1.1 & 13.24\\
E & 19 Jun 11 & 2455732.097 & 0.584 & 29~427 & 0.211 & 0.016 & 0.204 & 0.003 & 0.136 & 0.004 & 4.3 & 0.8 & 13.41\\
E & 22 Jun 11 & 2455735.130 & 0.598 & 74~032 & 0.213 & 0.007 & 0.205 & 0.002 & 0.135 & 0.003 & 4.1 & 0.4 & 13.41\\
E & 08 Jul 11 & 2455751.132 & 0.673 & 41~771 & 0.227 & 0.010 & 0.208 & 0.001 & 0.145 & 0.004 & 4.7 & 0.6 & 13.42\\
E & 13 Jul 11 & 2455756.134 & 0.696 & 52~216 & 0.225 & 0.009 & 0.208 & 0.003 & 0.145 & 0.005 & 4.2 & 0.5 & 13.35\\
E & 15 Jul 11 & 2455758.138 & 0.705 & 62~607 & 0.226 & 0.007 & 0.214 & 0.003 & 0.145 & 0.005 & 4.6 & 0.4 & 13.39\\
N & 25 Sep 11 & 2455829.535 & 0.037 & 50~501 & 0.205 & 0.005 & 0.222 & 0.002 & 0.157 & 0.002 & 6.7 & 0.5 & 13.39\\
N & 10 Oct 11 & 2455845.407 & 0.111 & 47~385 & 0.206 & 0.008 & 0.223 & 0.002 & 0.151 & 0.002 & 6.0 & 0.6 & 13.38\\
E & 30 Oct 11 & 2455865.444 & 0.204 & 54~215 & 0.183 & 0.006 & 0.208 & 0.005 & 0.141 & 0.002 & 3.7 & 0.5 & 13.38\\
N & 16 Nov 11 & 2455882.425 & 0.283 & 55~492 & 0.180 & 0.009 & 0.211 & 0.004 & 0.143 & 0.005 & 4.3 & 0.4 & 13.34\\
N & 27 Nov 11 & 2455893.309 & 0.334 & 60~720 & 0.191 & 0.006 & 0.216 & 0.006 & 0.147 & 0.005 & 2.7 & 0.4 & 13.35\\
N & 08 Dec 11 & 2455904.316 & 0.385 & 60~838 & 0.203 & 0.006 & 0.216 & 0.002 & 0.154 & 0.005 & 1.3 & 0.4 & 13.35\\
N & 09 Jan 12 & 2455936.252 & 0.534 & 49~042 & 0.209 & 0.005 & 0.216 & 0.002 & 0.159 & 0.005 & 4.0 & 0.5 & 13.25\\
N & 22 Jan 12 & 2455949.279 & 0.594 & 61~673 & 0.184 & 0.004 & 0.213 & 0.001 & 0.145 & 0.003 & 5.2 & 0.4 & 13.32\\
\\
\hline                                   
\end{tabular}
\end{center}
\end{table*}

The star $\beta$ Ceti (HD 4128, HR 188, HIP 3419) is one of the magnetic single late-type giants from the sample of Auri\`ere et al. (2009). It is a star of spectral class K0 III with $V = 2.04$ mag and $B-V = 1.02$ mag. Its value of $V-R = 0.72$ mag places it to the left of the coronal dividing line in the Hertzsprung-Russel diagram (hereafter HRD) where stars are supposed to have a chromosphere, a transition region, and a corona. That dividing line was proposed by Linsky \& Haisch (1979) near $V-R = 0.80$ mag and later confirmed by the study of Simon et al. (1982) with a larger sample of stars. In the solar neighborhood ($d \le 30$ pc), $\beta$ Ceti is the single giant star with the highest X-ray luminosity $\log L_{x} = 30.2$ erg/s (Maggio et al. 1998; H\"{u}nsch et al. 1996), which is calculated assuming a distance of $d = 29.5$ pc based on the Hipparcos parallax (van Leeuwen 2007). This high X-ray emission reaches a level comparable to that of Capella (a member of a long-period RS CVn association) and $\theta^{1}$ Tau (K0 III, in the Hyades open cluster) and reveals an extended corona. An atmospheric model of $\beta$ Ceti was simulated by Eriksson et al. (1983), which suggested there are coronal loops. Ayres et al. (2001) reported a series of striking coronal flare events observed with EUVE during a period of 34 days starting on 1 Aug 2000, confirming the strong activity of this single giant.

This paper, focusing on $\beta$ Ceti, is the third in a series of studies making use of the ZDI technique to investigate the magnetic topologies of cool giants. We follow here a first article dedicated to EK Eridani, which is likely to be a descendant of an Ap star (Auri\`ere et al. 2011), and a second one focused on the fast-rotating giant V390 Aur (Konstantinova-Antova et al. 2012). The purpose of this project is to investigate the magnetism of late-type active giants by determining how various stellar parameters (in particular the evolutionary stage) affect the magnetic field topologies and stellar activity. Other data sets collected for different targets of our sample will be published in forthcoming articles.

For this study, we have collected high-resolution spectropolarimetric data in the period June 2010 -- January 2012. In Section 2 we describe the observations and data reduction. Section 3 details the results of our measurements and data modeling. In Section 4, we derive the mass and discuss the evolutionary status of $\beta$ Ceti. Then, in Section 5, we propose possible interpretations of the reconstructed large-scale surface magnetic topology of $\beta$ Ceti and the variability of activity indicators. Finally, Section 6 contains our concluding remarks.

\section{Observations and data reduction}

Observational data were obtained with two twin fiber-fed echelle spectropolarimeters -- Narval (Auri\`ere 2003), which operates at the 2-m Bernard Lyot T\'elescope (TBL) at Pic du Midi Observatory, France, and ESPaDOnS (Donati et al. 2006 a), which operates at the 3.6-m Canada-France-Hawaii Telescope (CFHT) of Mauna Kea Observatory, Hawaii. In polarimetric mode, both have a spectral resolution of about 65\,000 and a nearly continuous spectrum coverage from the near-ultraviolet (at about 370 nm) to the near-infrared domain (at 1050 nm) in a single exposure, with 40 orders aligned on the CCD frame by two cross-disperser prisms. Stokes I (unpolarized light) and Stokes V (circular polarization) parameters are simultaneously obtained by four subexposures between which the retarders -- Fresnel rhombs -- are rotated in order to exchange the beams in the instrument and to reduce spurious polarization signatures (Semel et al. 1993).

Thirty-eight spectra have been collected for $\beta$ Ceti in the period June 2010 -- January 2012 (Table~\ref{table:1}). First, reduced spectra were extracted using the automatic reduction software LibreEsprit, developed for Narval and ESPaDOnS. A detailed description of the algorithm implemented in this software can be found in Donati et al. (1997). As a second step, the least squares deconvolution (LSD) multiline technique (Donati et al. 1997) was applied to all observations. This widely used cross-correlation technique enables averaging of several thousand absorption atomic lines recorded throughout the \'echelle spectrum, generating a single Stokes I and V line profile. We employed here a line mask created from Kurucz (1993) atmospheric models. The line mask was calculated for an effective temperature of $T_{eff} = 5000$ K, $\log g = 3.0$ and a microturbulence of 2 km/s, consistent with the literature data for $\beta$ Ceti (Thevenin 1998; Hekker \& Mel\'endez 2007; Massarotti et al. 2008), resulting in a total of about 12\,700 spectral lines (selecting lines with a depth greater that $0.1I_c$, where $I_c$ is the continuum level, and after removal of chromospheric lines and of spectral domains affected by telluring bands). By doing so, the signal-to-noise ratio (S/N) is increased to the point where weak polarized Zeeman signatures can be detected in all 38 spectra, thanks to a final S/N of up to 56\,000. Actually, we have two more spectra of $\beta$ Ceti taken in September 15, 2010, and November 20, 2010, but both were excluded from our analysis due to low S/N values; measurements extracted from these low-quality spectra are not included in the tables and figures.

As a first estimate of the magnetic field strength, we computed the longitudinal component of the magnetic field (B$_{l}$), using the first-order moment method (Rees \& Semel 1979; Donati et al. 1997; Wade et al. 2000 a, b). According to the multiline model, we assumed here that the LSD line profiles possess a mean wavelength of 597 nm and a Land\'e factor of 1.28. The first moment was computed between radial velocity boundaries set to encompass the whole velocity span of Stokes V signatures. For $\beta$ Ceti that integration window has velocity boundaries of $\pm$ 19 km/s around the line center of LSD profiles. The longitudinal magnetic field B$_{l}$ (expressed in gauss) was computed from both LSD Stokes I and V profiles with the following equation:

\begin{equation}
 B_{l} =-2.14\times10^{11}\frac{\int vV(v)\,dv}{\lambda gc \int [1-I(v)]\,dv}
\end{equation}
where $v$ (km/s) is the radial velocity in the stellar restframe, $\lambda$ (in nm) is the mean wavelength of all spectral features involved in the LSD process (597 nm in our case), $g$ the average Land\'e factor (here 1.28), and c (km/s) the light velocity in vacuum.

We also took advantage of several classical activity proxies showing up in the spectra. In an approach consistent with our previous papers, we measured the line depths for H$\alpha$ and for the CaII infrared triplet component at 854.2 nm regarding the normalized continuum ($R_{c}$). We estimated the core emission of the CaII K line through the intensity ratio between the line core and the continuum intensity at 395 nm. We measured the radial velocity (RV) of the LSD Stokes I line profiles by simply performing the $\chi^2$ adjustment of a Gaussian function on the line core and then taking the central velocity of the Gaussian as our RV value.

\section{Results}

\subsection{Longitudinal magnetic field, line activity indicators, and radial velocity}

   \begin{figure}[!hc]
      \centering
      \includegraphics[width=10cm, height=16cm]{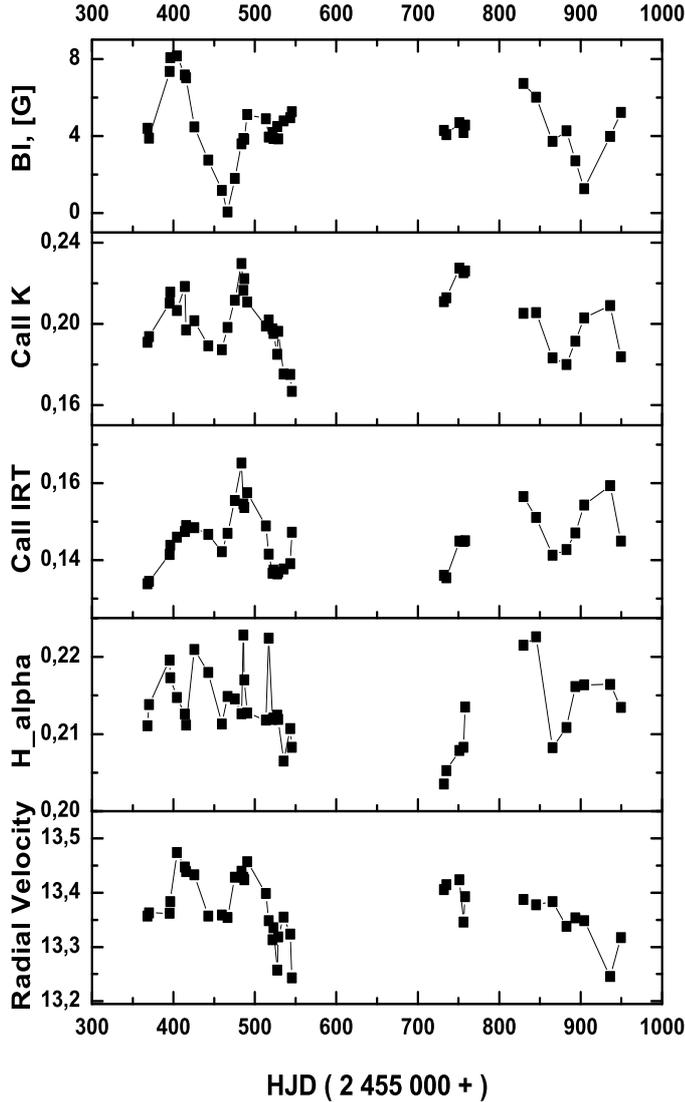}
      \caption{From top to bottom are presented the variations in B$_{l}$ and the activity indicators CaII K, CaII IR, H$\alpha$, and the radial velocity in the period June 2010 -- January 2012.}
   \label{fig:Bl}
   \end{figure}

All Stokes V profiles of $\beta$ Ceti have a simple shape with a peak-to-peak amplitude varying from $9.7\times10^{-5} I_c$ to $6.4\times10^{-4} I_c$ in 2010 and from $1.3\times10^{-4} I_c$ to $6\times10^{-4} I_c$ in 2011/2012. All line profiles feature a positive blue lobe and negative red lobe, which corresponds to a field of positive polarity. Accordingly, B$_{l}$ estimates stay with a positive sign for the whole observational period (see Table~\ref{table:1} and Fig.~\ref{fig:Bl}). Thanks to a typical error bar as low as 0.5~G, significant variations are observed in the interval 0.1 -- 8.2~G in 2010 and 1.3 -- 6.7~G in 2011/2012. The smaller amplitude of variations observed in 2011/2012 may be plausibly explained by the fact that the maximum values of B$_{l}$ have been missed in this dataset, due to the lack of observations between $HJD = 2455545$ and $HJD = 2455730$. We also noticed that B$_{l}$ displays almost sinusoidal temporal variations (Fig.~\ref{fig:Bl}) that would make the large-scale field distribution roughly consistent with a simple dipolar configuration (Landstreet \& Mathys 2000).

The chromospheric activity measurements derived from CaII K, CaII IRT, and H$\alpha$, as well as their uncertainties, are also listed in Table~\ref{table:1} and overplotted with the B$_{l}$ estimates in Fig.~\ref{fig:Bl}. Chromospheric emission follows the same general variations as B$_{l}$, especially for CaII K. This good agreement provides us with another hint that the surface magnetic geometry of $\beta$ Ceti is topologically rather simple. If B$_{l}$ provides us with a selective tracer of the largest scale magnetic structures (owing to the mutual cancellation of polarized signatures of nearby magnetic regions with opposite polarities), the chromospheric emission is sensitive to the field strength alone and is therefore unaffected by this spatial filtering effect. The good temporal agreement between B$_{l}$ and chromospheric proxies is therefore indicative of a lack of smaller scale magnetic regions outside rotational phases of maximum B$_{l}$, in striking difference to very active cool stars on the main sequence for which B$_{l}$ and chromospheric emission seem to be mostly uncorrelated (Morgenthaler et al. 2012).

Temporal variations in the radial velocity are shown in the lower right panel of Fig.~\ref{fig:Bl}, displaying values in the interval 13.24 -- 13.47 km/s (while the RV stability of Narval and ESPaDOnS was estimated by Moutou et al. (2007) to be of the order on 30 m/s), with relatively good correlation with other activity indicators. In early measurements, Frost (1923) reports $RV = 13.5$ km/s, while Buscombe \& Kennedy (1968) obtained a slightly higher value of $14.1 \pm 0.33$ km/s. A later estimate of $13.32 \pm 0.05$ km/s was proposed by Massarotti et al. (2008), in good agreement with our series of values.

\subsection{Zeeman Doppler Imaging}
\label{sec:ZDI}

We exploited the rotational modulation of Stokes V signatures to reconstruct the magnetic topology of the star by means of the Zeeman Doppler Imaging tomographic method (ZDI; Semel 1989; Donati \& Brown 1997; Donati et al. 2006 b). We used a recent implementation of this algorithm where the surface vectorial magnetic field is projected onto a spherical harmonics frame, allowing us to easily distinguish between the poloidal and toroidal components of the surface magnetic geometry (Donati et al. 2006 b). This method performs iterative adjustment of the observed time series of LSD polarized profiles by a simulated set of Stokes V profiles computed for an identical sequence of rotational phases. The synthetic Stokes profiles are calculated from an artificial star whose surface is divided into a grid of 2\,000 rectangular pixels of roughly similar area. Each surface pixel is associated with a local Stokes I and V profile.

In this simple model, the local synthetic Stokes I line profile is assumed to possess a Gaussian shape, with a depth and width adjusted to achieve the best fit between synthetic and observed line profiles. Assuming a given magnetic field strength and orientation for each pixel, local Stokes V profiles are calculated under the weak-field assumption (Morin et al. 2008; Petit et al. 2010; Morgenthaler et al. 2012). The rotation period, inclination angle, and $v \sin i$ adopted for our model are discussed below. The linear limb darkening coefficient is set to 0.75, in agreement with Claret \& Bloemen (2011). We assume here that the surface magnetic field of the star contains both a poloidal and toroidal field component (which is consistent with the previous observation of a number of cool active stars, e.g. Konstantinova-Antova et al. 2008). We limited the spherical harmonics expansion to $l \le 10$ since no improvement is found for the fits between modeled and observed LSD profiles for higher values of $l$.

\begin{figure}[!htc]
    \centering
    \includegraphics[width=10cm, height=8cm]{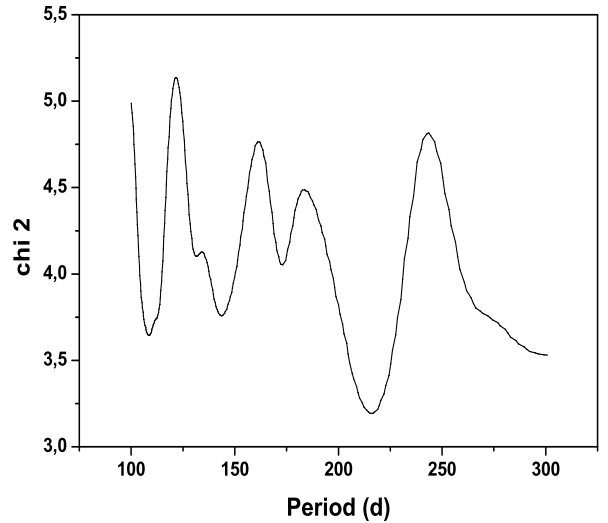}
     \caption{Periodogram obtained using the whole dataset June 2010 -- January 2012.}
   \label{fig:period}
   \end{figure}

The first step towards the ZDI reconstruction of the magnetic topology of $\beta$ Ceti is to determine the rotational period of the star. We conducted the period search from the full dataset (June 2010 -- January 2012), following the approach of Petit et al. (2002). Two hundred ZDI models were computed (following the method detailed in Section 2), assuming a different rotation period of the star for each model, with a uniform sampling of rotation periods in the 100- to 300-day interval. Forcing a constant information content (i.e. a same average magnetic field strength) in all 200 magnetic topologies, we compared them to the observed Stokes V profiles and analyzed the $\chi^2$ variations over the period span (Fig.~\ref{fig:period}). Our best magnetic model, identified by the lowest $\chi^2$ value, suggests a possible period of 215 days, which we use for the final magnetic map. A slightly shorter period of 199 days was derived by Jordan \& Montesinos (1991) using the value of $v \sin i = 3$ km/s given by Gray (1982).

We applied the same maximum-entropy method to look for surface differential rotation (Petit et al. 2002), but did not obtain any conclusive result. The rotational phase of every Stokes V profile (Table~\ref{table:1}) is computed according to the rotation period of 215~d, with a phase origin at $HJD = 2454101.5$. All data are phased according to the following ephemeris:

\begin{equation}
 HJD=2454101.5 + 215\, \phi
\end{equation}
where HJD is the heliocentric Julian date of the observations, and $\phi$ is the rotational cycle.

It was found from the literature that $\beta$ Ceti has a rotational velocity in the interval $v \sin i = 3 - 4$ km/s ($3.3 \pm 0.8$ km/s from Smith \& Dominy (1979); $4 \pm 1$ km/s from Fekel (1997); $3 \pm 1$ km/s from Carney et al. (2008)). Varying the value of $v \sin i$ over this range in our ZDI models did not significantly affect the resulting maps and $\chi^2$ values. In the rest of this paper, we therefore chose an intermediate value $v \sin i = 3.5$ km/s. The inclination angle  $i$ of the stellar spin axis is the last ZDI input parameter. A value of $i = 60^\circ$ was proposed by Sanz-Forcada et al. (2002), following parameters from Gray (1989). Combining our value of $v \sin i$, our estimate of the stellar radius (see Section 4) and the rotational period derived from ZDI, we also adopt here $i = 60^\circ$ in the tomographic models.

\begin{figure*}
    \centering
    \includegraphics[width=6cm, height=10cm]{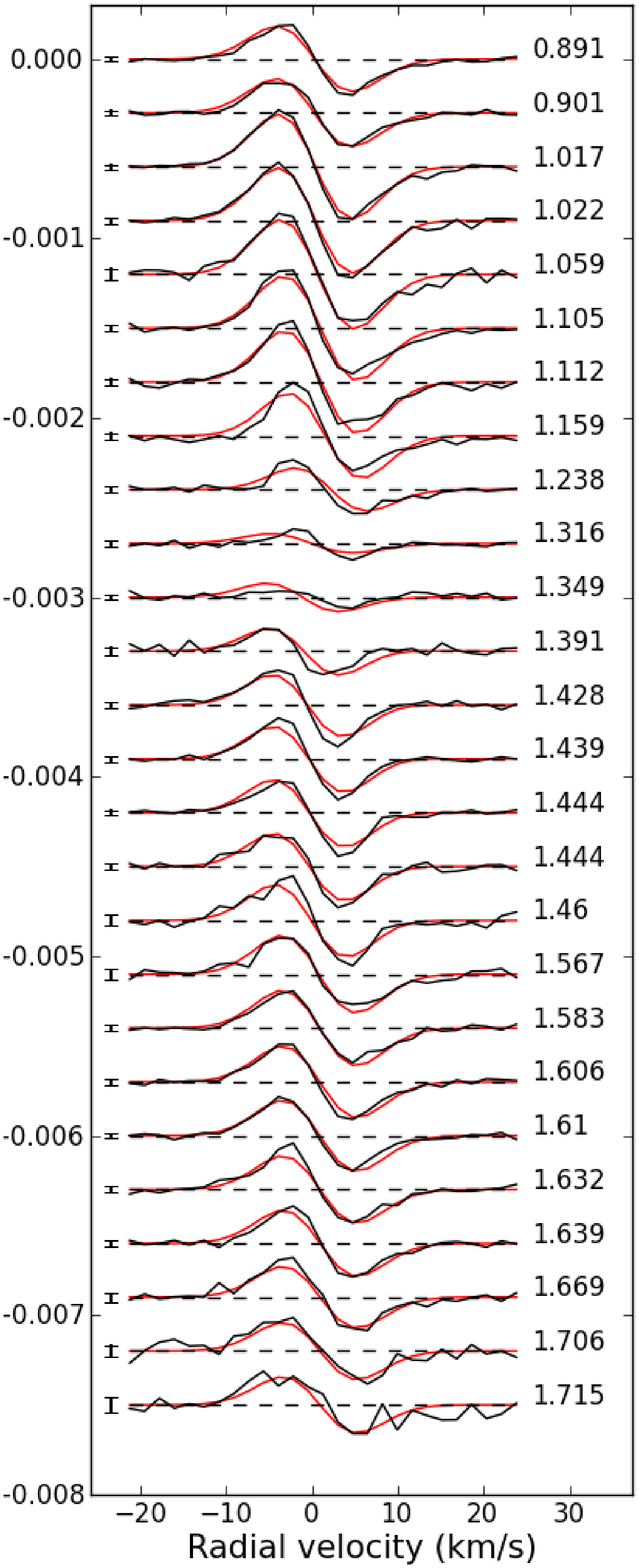}
    \includegraphics[width=10cm, height=10cm]{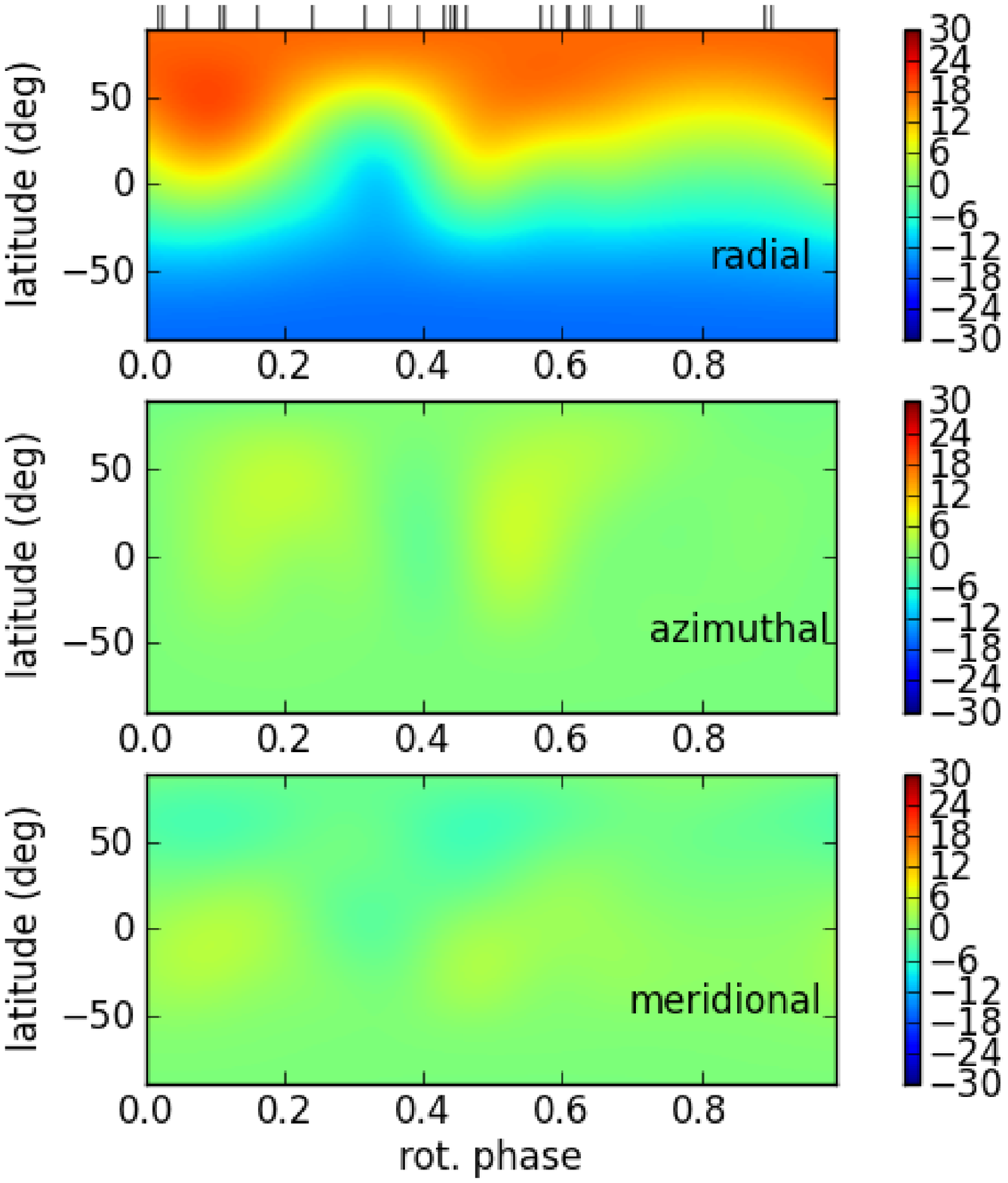}
     \caption{$\beta$ Ceti in the period June 2010 -- December 2010. Left panel: Normalized Stokes V profiles - observed profiles (black lines); synthetic profiles (red lines); zero level (dashed lines). All profiles are shifted vertically for display purposes. The rotational phases of observations are indicated in the right part of the plot and the error bars are on the left of each profile. Right panel: Magnetic map of $\beta$ Ceti. The three panels illustrate the field components in spherical coordinates (from top to bottom -- radial, azimuthal, meridional). The magnetic field strength is expressed in gauss. The vertical ticks on top of the radial map show the phases of observations.}
   \label{fig:zdimap1}
   \end{figure*}

\begin{figure*}
    \centering
    \includegraphics[width=6cm, height=10cm]{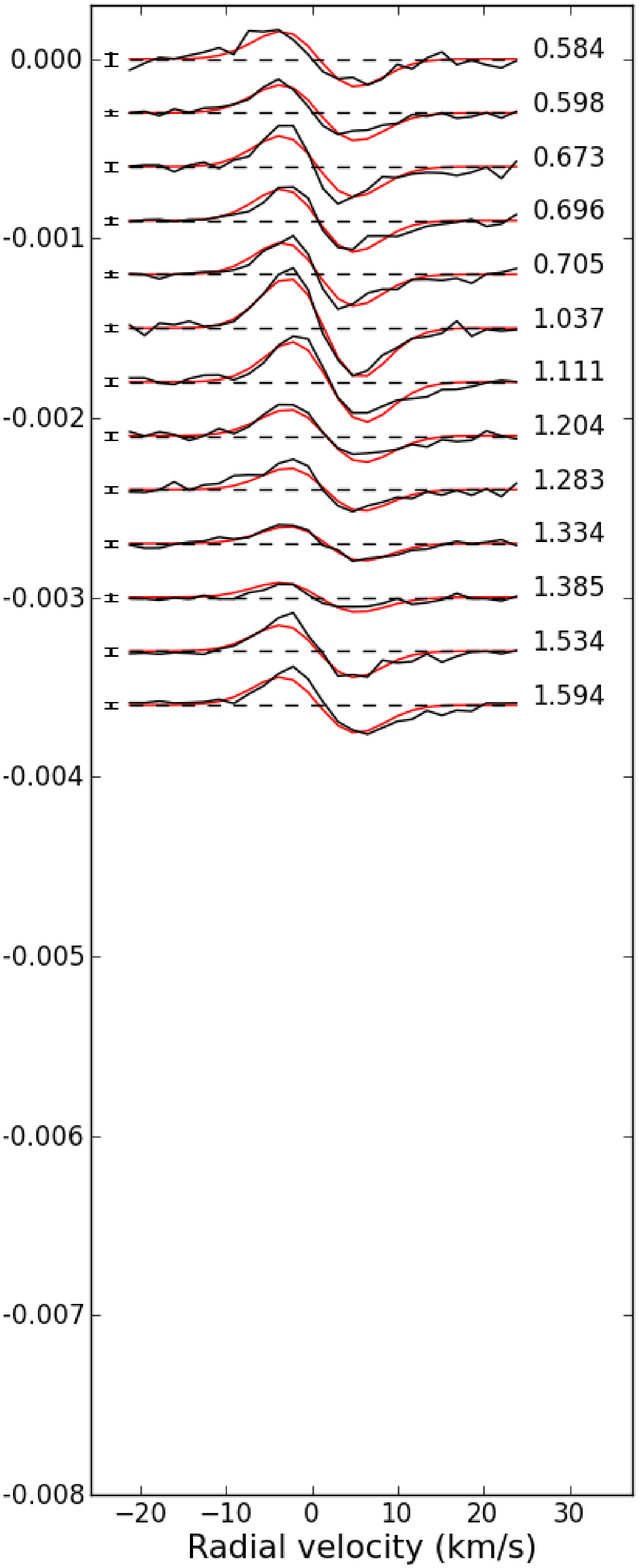}
    \includegraphics[width=10cm, height=10cm]{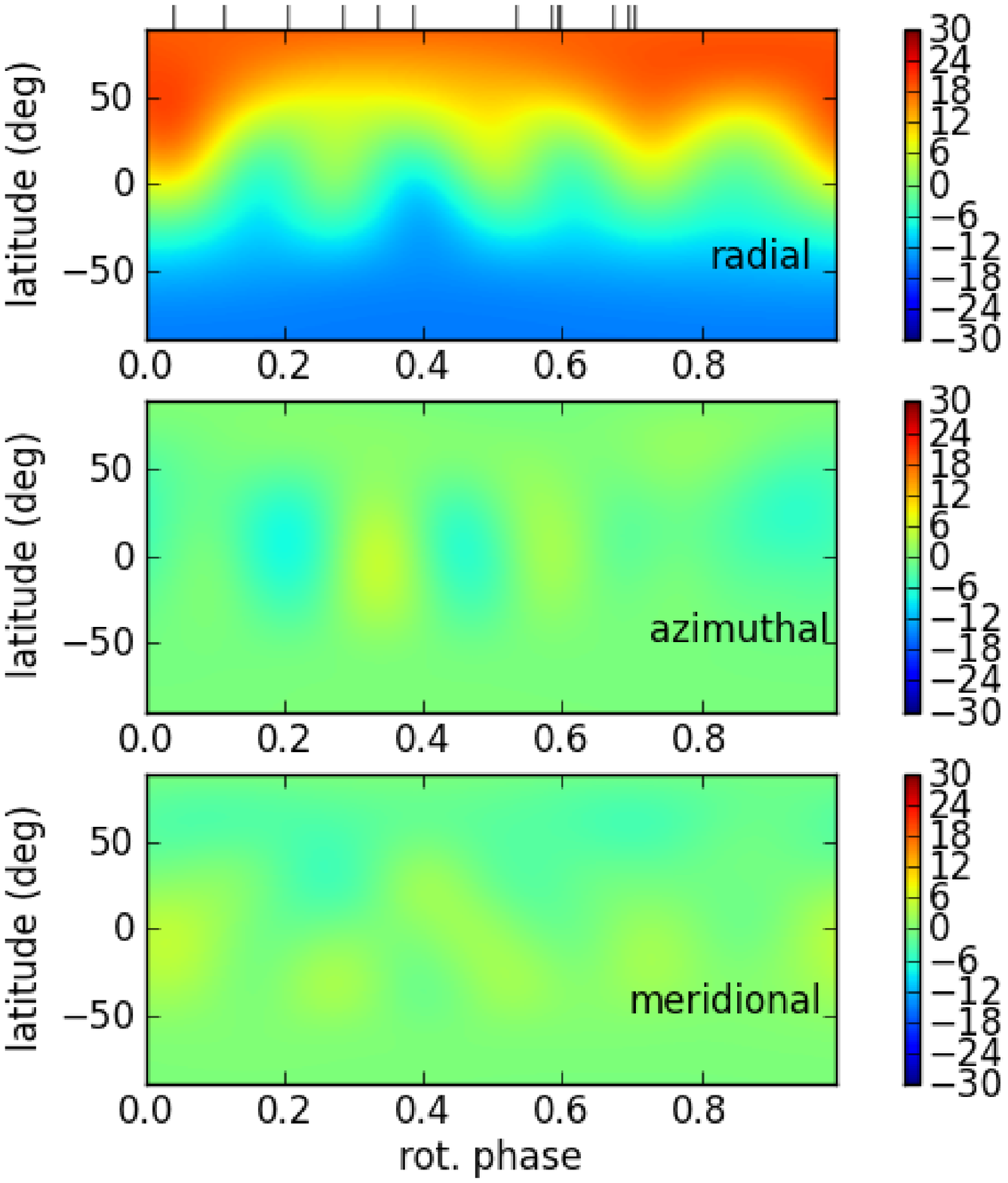}
     \caption{Same as Fig.~\ref{fig:zdimap1}, for the period June 2011 -- January 2012.}
   \label{fig:zdimap2}
   \end{figure*}

In this study we present two magnetic maps for $\beta$ Ceti, choosing to split our time series into two subsets (one for June 2010 -- December 2010 and the second one for June 2011 -- January 2012, see Figs.~\ref{fig:zdimap1} and \ref{fig:zdimap2}, respectively). This strategy is meant to highlight possible changes in the large-scale surface field between the two epochs (if any). The subsets obtained for 2010 and 2011/2012 cover similar time spans and contain 25 and 13 spectra, respectively. According to the rotation period of 215 days, both subsets provide us with sufficiently dense phase sampling to reconstruct a magnetic map, in spite of a lack of observations in the second dataset at phases higher than 0.7.

Both ZDI maps (illustrated in Figs.~\ref{fig:zdimap1} and \ref{fig:zdimap2}) correspond to magnetic models with $\chi^2 = 2.1$. The Stokes V data are therefore not fit at the noise level, mainly because our simplified magnetic model does not offer the possibility of generating asymmetric Zeeman signatures, while the LSD profiles of $\beta$ Ceti clearly display such asymmetry (with blue lobes generally deeper than red lobes). Such abnormal Stokes V profile shapes have been previously observed on several active stars (e.g. Auri\`ere et al. 2008) and most likely originate in vertical gradients of velocity and magnetic field (Lopez Ariste 2002). Previous attempts to get a better fit by implementing an ad hoc asymmetry in the model (Petit et al. 2005) did not significantly modify the resulting magnetic topology.

\begin{table*}[!htc]
\centering
\begin{center}
\caption{Magnetic characteristics for the two magnetic maps of $\beta$ Ceti. The fifth to ninth columns list the fraction of the large-scale magnetic energy reconstructed in the poloidal field component, the fraction of the poloidal magnetic energy stored in the dipolar ($l=1$), quadrupolar ($l=2$) and octopolar ($l=3$) components, and the fraction of the energy stored in the axisymmetric component ($m=0$).}  
\label{table:dipoledata}      
\centering                          
\begin{tabular}{c c c c c c c c c}        
\hline               
      & rotational phase & latitude of    & magnetic & pol.     & dipole   & quad.    & oct.     & axi.     \\
Epoch & of maximum       & maximum        & dipole   & comp.    & comp.    & comp.    & comp.    & comp.    \\
      & field strength   & field strength & strength & (\% tot) & (\% pol) & (\% pol) & (\% pol) & (\% tot) \\
\hline                        
2010        & 0.1 & $52^\circ$ & 20.5 G & 96.7 & 83.2 & 20.8 & 6.2 & 77.1 \\
2011/2012   & 0.0 & $50^\circ$ & 20.6 G & 96.5 & 85.4 & 11.3 & 4.0 & 74.4 \\
\hline                                   
\end{tabular}
\\
\end{center}
\end{table*}

Most properties of the two magnetic maps are very alike. At both epochs, the magnetic geometry is dominated by the poloidal component of the magnetic field, which contains about 96\% of the reconstructed magnetic energy (Table~\ref{table:dipoledata}). Most magnetic energy of the poloidal component is also stored in spherical harmonics modes with $\ell = 1$, with 83\% and 85\% (for the first and the second maps, respectively) of the reconstructed poloidal energy. This dominant contribution of low-order modes unveils a very simple topology, dominated by a dipolar configuration (the contribution of which is especially visible in the radial field component). The dipole is almost aligned on the spin axis, leading to a majority of the magnetic energy in modes with $m = 0$. The visible pole of the dipole features a positive field polarity and a polar field strength close to 20~G (see Table~\ref{table:dipoledata}), in agreement with positive B$_{l}$ values recorded for $\beta$ Ceti. We also report in Table~\ref{table:dipoledata} a limited shift in rotational phase of the magnetic pole between the two maps. This apparent phase drift may be linked to the phase gap in the 2011/2012 dataset close to the positive pole (at phases 0.7 and above), which may have induced a less accurate longitudinal positioning of the pole.

To check the good consistency of both magnetic topologies again, we computed another ZDI map (not shown in this paper) combining all the data from June 2010 to January 2012. This global magnetic model is obtained with a $\chi^2$ of 2.6, slightly larger than the $\chi^2$ of the two separate maps. If the slightly worse Stokes V fit suggests a (limited) temporal evolution of the field structure, the resulting topology still agrees with both topologies obtained from individual subsets, which indicates that the temporal changes have left the dipole mostly untouched and are more likely related to smaller scale features of the magnetic geometry.

\section{Mass, evolutionary status, and surface abundances of $\beta$ Ceti}

\begin{figure}[!hc]
    \centering
    \includegraphics[width=8cm, height=8cm]{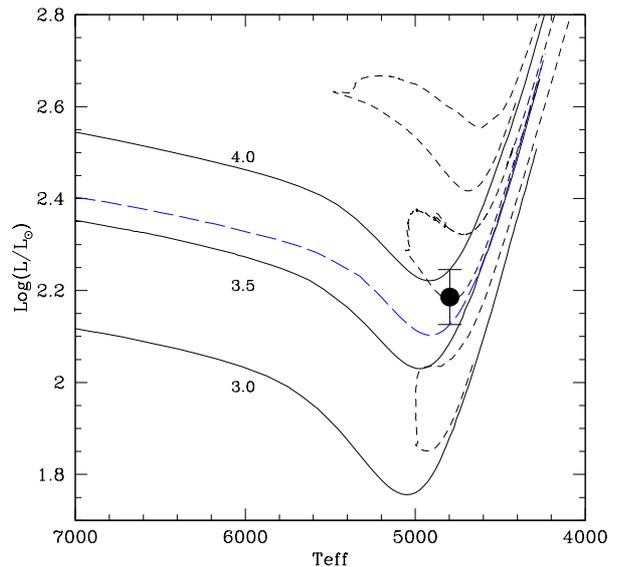}
     \caption{Position of $\beta$ Ceti in the Hertzsprung-Russel diagram using the effective temperature by Massarotti et al. (2008) and the luminosity derived from Hipparcos data (see text for details). Standard evolution tracks at solar metallicity by Charbonnel \& Lagarde (2010) are shown in black for different initial masses as indicated by labels, with solid and short dashed lines indicating the phases of evolution before and after ignition of central helium-burning, respectively. The blue long dashed line is for the 3.5~M$_{\odot}$ rotating model (see text).}
   \label{fig:hrdCC}
   \end{figure}

The position of $\beta$ Ceti in the Hertzsprung-Russell diagram is shown in Fig.~\ref{fig:hrdCC}. We adopt the effective temperature of 4797~K from Massarotti et al. (2008) and the luminosity is computed using the stellar parallax from the New Reduction Hipparcos catalog by van Leeuwen (2007), the V magnitude from 1997 Hipparcos catalog, and the bolometric correction from Flower (1996; BC = --0.403); the error bar on luminosity reflects only the uncertainty on the parallax. The corresponding stellar radius obtained with the Stefan-Boltzmann law is 18~R$_{\odot}$.
\\

We infer a mass of 3.5~M$_{\odot}$ based on the standard evolution tracks at solar metallicity from Lagarde et al. (2012), which are plotted in Fig.~\ref{fig:hrdCC}. According to our models, the main sequence progenitor of $\beta$ Ceti was a late B-type star. Maggio et al. (1998) give an A-type star for a progenitor. Our values for the radius and  mass are compatible with those found in the literature (15 -- 17~$R_{\sun}$ and 2.8 -- 3.5~$M_{\sun}$; Jordan \& Montesinos 1991; Maggio et al. 1998; Schr\"{o}der et al. 1998; Gondoin 1999; Allende Prieto \& Lambert 1999; Massarotti et al. 2008; Berio et al. 2011).

As seen from its position in the HRD in Fig.~\ref{fig:hrdCC}, $\beta$ Ceti could presently either be at the base of the red giant branch or be undergoing core helium-burning. The latest option is actually strongly favored because of the much longer duration of the central helium-burning phase. In addition, the carbon isotopic ratio $^{12}$C/$^{13}$C $=  19 \pm 2$ (Luck \& Challener 1995; Tomkin et al. 1975) and the Li abundance N(Li) $= 0.01$ (Luck \& Challener, 1995) derived for $\beta$ Ceti indicate that this star has already fully undergone the first dredge-up. Indeed as can be seen in Table~\ref{table:lithiumcarbon} where we give these quantities for standard and rotating 3.5~M$_{\odot}$ models at the luminosity and effective temperature of $\beta$ Ceti both for the first RGB ascent and helium-burning options, the observational data are in excellent agreement with the theoretical post-dredge-up predictions for the surface carbon isotopic ratio and lithium abundance of the 3.5~M$_{\odot}$, Z$_{\odot}$ rotating model shown in Fig.~\ref{fig:hrdCC}. Clearly, first ascent is excluded by the observed carbon isotopic ratio since at this position the first dredge-up is still going on, while it is fully completed when the star undergoes central helium-burning (3.5~M$_{\odot}$ star does not undergo thermohaline mixing on the RGB; see Charbonnel \& Lagarde 2010 for more details). Additionally, the low lithium abundance for such an intermediate-mass star is a robust signature of rotation-induced mixing when the star was on the main sequence (see e.g. Palacios et al. 2003).

The 3.5~M$_{\odot}$ rotating model was computed with the same input physics and assumptions as Lagarde et al. (2012). In particular, initial velocity on the zero age main sequence is chosen equal to 45$\%$ of the critical velocity, which corresponds to the mean value of the observed velocity distribution for cluster and field B type stars (Huang et al. 2010). At the location of $\beta$ Ceti (helium-burning), the theoretical surface rotation velocity of the model is 6~km/s, in agreement with the observed $v$sin$i$ for this star, although it corresponds to a rotational period of 160 days, i.e. slightly lower than the value we derived through ZDI in Section 3.2.

\begin{table}[hccccc]
\centering
\begin{center}
\caption{Theoretical predictions for surface Li abundance and carbon isotopic ratio at the luminosity and effective temperature of $\beta$ Ceti on the first ascent of the red giant branch and on the central helium-burning phase in the 3.5~M$_{\odot}$ standard and rotating models for two initial rotation velocities (50 and 140 km.s$^{-1}$). These numbers have to be compared with the observational values for $\beta$ Ceti, namely N(Li)= 0.01 and $^{12}$C/$^{13}$C = 19$\pm$2 (see text for references and details).}  
\label{table:lithiumcarbon}      
\centering                          
\begin{tabular}{c c c c c}        
\hline               
 &  N(Li) & N(Li) & $^{12}$C/$^{13}$C  & $^{12}$C/$^{13}$C  \\
& RGB & He-burning & RGB & He-burning \\
\hline                        
standard      & 1.75 & 1.27 & 67 & 20.6 \\
rotation (50)   & 1.58 & 1.12 & 61 & 20.6 \\
rotation (140) & 0.53 & 0.007 & 38 & 18.32 \\
\hline                                   
\end{tabular}
\\
\end{center}
\end{table}

\section{Discussion}

The detection of a magnetic field on $\beta$ Ceti (with a longitudinal strength of about 8 gauss) was first reported by Auri\`ere et al. (2009). We present here our extended work for this star. Using spectropolarimetric data, we derived two magnetic maps of $\beta$ Ceti. We reconstructed the surface magnetic field topology of the star for two sets of observations (June 2010 -- December 2010 and June 2011 -- January 2012), and our model suggests that the photospheric magnetic geometry is mostly stable and dominated by a simple dipole. Smaller scale features of the surface field topology are simply out of reach of the ZDI technique, because of its low spatial resolution at small $v \sin i$ values (e.g. Petit et al. 2005). The sharp line profiles also prevented us from applying the classical Doppler Imaging method to our set of Stokes I LSD profiles for reconstructing the brightness map associated with the magnetic topology, as previously done for giants rotating faster, e.g. Petit et al. (2004). In such a situation, the analysis of activity indicators is a useful complementary piece of information to partially reveal the presence and phase distribution of smaller magnetic elements. For $\beta$ Ceti, it is clear from Fig.~\ref{fig:Bl} that the variations in the longitudinal magnetic field B$_{l}$ and chromospheric indicators correlate rather well, giving further support to the hypothesis of a field distribution limited to a very simple geometry.

The chromospheric activity of $\beta$ Ceti is at the level observed on giant stars with $P_{rot} \approx 80$ d (Young et al. 1989), while the B$_{l}$ maximum value of 8 G is typical of G and K giants with periods shorter than about 100 days (Auri\`ere et al., in preparation). With a much longer estimated period of 215~d, $\beta$ Ceti exhibits an abnormally high activity level, at least if we bear in mind the simple picture of a global dynamo efficiency that increases with the surface rotation rate. We also note that the simple field topology reconstructed from our observations is clearly different from the field geometry previously obtained for active giants rotating faster because of its simple structure and because of the lack of any significant toroidal component (Petit et al. 2004, Konstantinova et al. 2008). Another clear difference with more rapid rotators is the absence of any measurable surface differential rotation. The shear levels reported for other evolved objects are able to deeply modify field topologies within a few weeks (e.g. Petit et al. 2004), while $\beta$ Ceti seems to keep a mostly constant field geometry over several years.

Since the magnetic behavior of $\beta$ Ceti seems to be partly at odds with the global dynamo framework, we propose to explore the alternate option of a fossil magnetic field that may be inherited from the stable field of an Ap star, considering that the mass estimated for $\beta$ Ceti (3.5 M$_{\odot}$) corresponds to a late-B/early-A star on the main sequence. This hypothesis was already proposed for EK Eridani (Auri\`ere et al. 2008; 2011), another abnormally active cool giant sharing with $\beta$ Ceti a magnetic topology almost purely poloidal and dominated by a dipole (in the relevant inclination range). Assuming that this fossil field is buried in the radiative zone, one may question if it could permeate the deep convective zone of a red giant, up to the photosphere. A similar situation has been simulated in the case of a 1~$M_{\sun}$ on the main sequence, for which the convective envelope occupies about 30\% of the radius. Strugarek et al. (2011 a; 2011 b) show that for an aligned or an inclined magnetic dipole, the magnetic field is able to expand from the radiative interior to the convective envelope.

The simple field topology of $\beta$ Ceti is reminiscent of the mainly dipolar field configurations observed on main-sequence Ap/Bp stars (see e.g. L\"{u}ftinger et al. 2010). Following the fossil field hypothesis further, we can derive a rough estimate of the surface field strength of the main sequence progenitor of $\beta$ Ceti. From our 3.5~M$_{\odot}$ model, we find that the radius of $\beta$ Ceti must have been $R(zams) = 2.01$~$R_{\sun}$ when the star arrived on the zero age main sequence. Assuming that the field measured for $\beta$ Ceti simply differs from the main-sequence field by the dilution effect of its larger radius, we can use the simple formula of Stepien (1993) for estimating the magnetic dipole strength of its main sequence parent star -- $B(MS) = B [ R / R(MS) ]^2$, where B and R are the present dipole strength and radius of the star and B(MS) and R(MS) are the dipole strength and radius on the main sequence. In the hypothesis of conservation of magnetic flux, we derive a magnetic dipole strength of $B(MS) \approx 1650$ G on the main sequence, which is consistent with typical field strength measured on Ap stars (Landstreet \& Mathys 2000; Auri\`ere et al. 2007).

In the assumption of $\beta$ Ceti being a descendant of an Ap star, we finally computed an evolutionary model of slow rotator with magnetic braking after the turnoff. For $V_{init} = 50$ km/s, we did not find any difference for Li abundance or $^{12}$C/$^{13}$C using models with and without magnetic braking after the turnoff. The numbers presented in Table~\ref{table:lithiumcarbon} are therefore still relevant under this assumption, leaving our conclusions unchanged about the evolutionary state of $\beta$ Ceti.

\section{Conclusions}

We collected spectropolarimetric data for the cool giant $\beta$ Ceti in the period June 2010 -- January 2012. The temporal modulation of polarized signatures is consistent with a rotation period of 215~d. Using the spectropolarimetric dataset, we were able to reconstruct two magnetic maps (taken one year apart) of the surface magnetic field topology of the star employing the Zeeman Doppler Imaging technique. Both magnetic maps display very similar magnetic geometries, with a global magnetic field dominated by a dipolar configuration, with a dipole axis almost aligned on the spin axis. The polar field strength is close to 20~G for both maps.

The longitudinal magnetic field B$_{l}$ has significant variations in the interval 0.1 -- 8.2~G and stays with a positive polarity for the whole observational period. Sinusoidal variations of B$_{l}$ are consistent with a large-scale dipole field configuration. The behavior of the line activity indicators H$\alpha$, CaII K, CaII IR, and the radial velocity correlate rather well with the B$_{l}$ variations, giving further support to the dipolar model produced by ZDI.

Based on a comparison with recent stellar models (computed with the same assumptions and input physics as in Charbonnel \& Lagarde 2010), we derive a mass of 3.5~M$_{\odot}$ for $\beta$ Ceti and propose that this star is already in the central helium-burning phase rather than starting the first ascent of the red giant branch. The evolutionary status is also supported by comparison between observed and theoretical values for the lithium abundance and carbon isotopic ratio at the stellar surface. Based on our evolutionary models, we derive a convective turnover time equal to $\tau_{c} = 171$ days, and we estimate the Rossby number of $\beta$ Ceti to be $Ro = 1.26$, suggesting that a large-scale dynamo alone may not be able to account for the high activity of the star.

We derived four useful clues here to better understand the magnetic nature of $\beta$ Ceti with (a) a mass of 3.5~M$_{\odot}$, which is consistent with the typical masses of chemically-peculiar stars; (b) a long rotation period of 215~d that seems unable to account for the high activity level recorded for this object through dynamo action alone; (c) a simple and stable magnetic topology consistent with dominantly dipolar field geometries observed in Ap/Bp stars; (d) a polar field strength of about 20~G which, when back-extrapolated to the main sequence, is consistent with typical field strengths of Ap/Bp stars. Considering all this information together, we propose that the magnetism of $\beta$ Ceti may be (at least partly) of fossil origin and inherited from a main sequence Ap/Bp star.

The present study, together with the studies of EK Eri (Auri\`ere et al. 2008, 2011) and 14 Ceti (Auri\`ere et al. 2012), may give a first view on the magnetic field evolution of an Ap star at different stages after the main sequence, Hertzsprung gap, base of the red giant branch, and the helium-burning phase.

\begin{acknowledgements}
We thank the TBL and CFHT teams for providing service observing with Narval and ESPaDOnS. The observations in 2010 with Narval were funded under Bulgarian NSF grant DSAB 02/3/2010. R.K.-A., S.Ts., and R.B. acknowledge partial financial support under NSF contract DO 02-85. R.B. also acknowledges support under the RILA/EGIDE exchange program (contract RILA 01/14). G.A.W. acknowledges support from the Natural Sciences and Engineering Research Council of Canada (NSERC). S.Ts. is thankful for the possibility to work for 3 months in 2011 in IRAP, Tarbes, France, under the Erasmus program and the contract DMU 03-87 for the possibility to work for 2 weeks in 2012 in IRAP, Toulouse, France. C.C. and T.D acknowledge support from the Swiss National Foundation (FNS) and the French Programme National de Physique Stellaire (PNPS) of CNRS/INSU.
\end{acknowledgements}

\end{document}